# Vibrational Anharmonicity Assisted Phase Transitions in Perovskite Oxides under Terahertz Irradiation

Cong Zhou, Jian Zhou[*]

*Center for Alloy Innovation and Design, State Key Laboratory for Mechanical Behavior of Materials, Xi'an Jiaotong University, Xi'an, 710049, China*

Abstract

Despite extensive research interests in perovskite oxides, low energy consumption, non-destructive and maneuverable methods for phase transition in perovskite oxides are still the under its exploration, and the underlying mechanisms remain ambiguous. Here, optical susceptibility including electronic and anharmonic phononic contributions is used to evaluate Gibbs free energy variations of $PbTiO_3$ and $BaTiO_3$ under terahertz irradiation. This corresponds to an off-resonant light-controlled phase transition, rather than the resonant approaches that excites hot carriers over electronic band or infrared-active vibrations in the phonon band. We show that intermediate terahertz light can trigger polarization change between ferroelectric orientation variants of $PbTiO_3$ at room temperature 300 K. Furthermore, the phase transition from low symmetric ferroelectric phase to high symmetric paraelectric structure in $PbTiO_3$ can be driven by changing the direction and intensity of the incident light under the same conditions. Similar results are observed in $BaTiO_3$. In detail, phonon spectrum and optical susceptibility are obviously modified and show temperature dependence, in which we show the significant effects of anharmonic vibration. In order to show its nonlinear optical nature, we perform an alternating electric field dressed *ab initio* molecular dynamics simulation, which maps the Raman-active phonon excitation under off-resonant terahertz irradiation.

---

[*] Email: jianzhou@xjtu.edu.cn

# I. Introduction

Since the discovery of ferroelectric (FE) materials over a century ago, they have possessed exotic applications such as nonvolatile electronics with fast switching speed,[1] light-electrical-mechanical energy conversion,[2] non-destructive readout,[3] capability of high-density integration, ultrafast data read/write kinetics, etc.[4] Until now, the most widely studied ferroelectric material type is perovskite oxides, which exhibits polymorphic lattice structures under different environmental condition, and could show ultrafast displacive phase transitions among them. Numerous studies have demonstrated that phase transition in perovskite oxides can be triggered via mechanical,[5-7] thermal,[8-10] and electrical approaches,[11] while recent experimental advances have witnessed such phase transitions in titanates under below bandgap light illumination.[12, 13] Note that the atomic diffusionless motion in perovskites during collective phase transitions corresponds to the Γ-mode vibrations in the momentum space, in accordance with the long wavelength feature of the optical field, which may largely reduce energy consumption and enhance its efficiency during phase transition. Theoretical computations have shown that electron excitation between the valence and conduction bands could induce phase transformations in perovskite oxides, via strong coupling between phonon vibration and light excited carriers.[14] In spite of these efforts, fundamental mechanisms of photo-induced structural change still await further exploration and understanding.

In this work, we use first-principles density functional theory (DFT) calculations including anharmonic vibrational effects to show that controllable phase transitions in perovskite oxides could occur under off-resonant THz light irradiation. The THz optics has been considered as an advanced technique owing to its noncontacting, transparency, non-destructive, and athermic nature.[15-18] Hence, it holds great potential for various promising applications, including non-destructive check,[19] food quality scan,[20] security check,[21] long-distance communication,[22] and modern electronic devices.[23, 24] We use two prototypical perovskite oxides, namely, lead titanate and barium titanate, to illustrate phase transitions under THz irradiation. The optical responses are evaluated through both electron and phonon contributions. We include anharmonic vibrational effects in the phonon dispersion, which could incorporate the thermal effects that have been largely ignored in conventional density functional perturbation approaches. Through thermodynamic evaluations, we suggest two possible phase transition paths under intermediate THz illumination. One is between two FE orientation variants (the Goldstone mode in a Mexican-hat potential profile), and the other is between the low symmetric FE phase and high symmetric paraelectric (PE) structure (the Higgs mode). They would occur depending upon the THz polarization and intensity. Here, THz frequency is chosen to be away from the infrared (IR) active phonon mode at Γ in the first Brillouin zone (BZ). It corresponds to an off-resonant light-matter interaction, rather than generating electron-hole pairs or IR-active mode, ensuring its athermic nature. Hence, we predict an alternative approach from the previously discussed schemes[25-27] that an IR-active vibration generates the phase transition responsible Raman modes, involving multiple (three or four) phonon scattering. We perform alternating electric field assisted *ab initio* molecular dynamics simulation to explicitly show that the Raman-active mode can be excited through this off-resonant light irradiation, consistency its being a nonlinear optical process.

**II. Results**

*A. Geometric, electronic and phonon properties of PbTiO$_3$*

We use lead titanate (PbTiO$_3$) to start our discussion, which exhibits two typical phases (**Figures 1a and 1b**). The high symmetric PE phase shows a cubic structure (space group of $Pm\bar{3}m$, no. 221), which usually appears under high temperature (above 766 K) in the conventional phase diagram. Our density functional theory (DFT, see Methods section) calculations give its lattice parameters to be $a = b = c = 3.92$ Å. Below the critical temperature, PbTiO$_3$ undergoes an inversion symmetry breaking phase transition, resulting in a FE tetragonal structure (space group of *P4mm*, no. 99). The polar is parallel to the principal axes, namely, ⟨100⟩ directions of the cubic phase (all together six equivalent directions). The lattice constant is slightly elongated along the polar direction (relaxed to be 4.09 Å), and the other two lattice parameters are equally to be 3.89 Å. These results agree well with the experimental measurements.[28] Note that the polarization direction could distinguish different FE phases with symmetrically equivalent states (e.g., orthogonal FE$_1$ and FE$_2$ orientational variants), separated by an energy barrier. **Figure 1c** schematically plots the potential energy surface as a function of electric polarization $P_z$ and $P_x$. Due to the $C_4$ rotation, there are four minima in the potential energy surface. The potential energy landscape of phase transitions between FE and PE is associated with longitudinal changes of polarization, similar as a Higgs mode.[29] The energy biasing between two different FE orientational variants (FE$_1$ and FE$_2$) corresponds to transverse changes of polarization, which belongs to a Goldstone mode.[29]

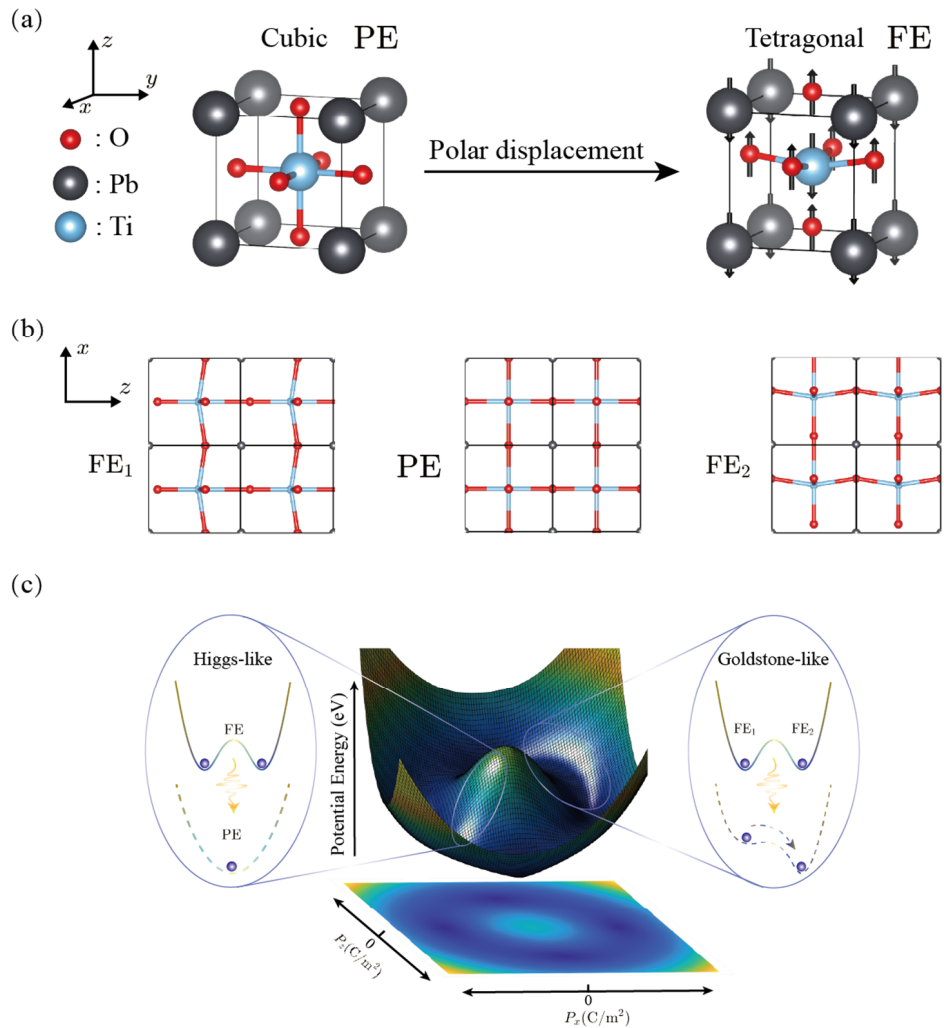

**Figure 1.** Schematic illustration of geometric phase transition in PbTiO$_3$. (a) Crystal

structures of the cubic paraelectric and tetragonal ferroelectric unit cells. The polar displacement is along the *z* axis. (b) The atomic structures of PbTiO$_3$ in FE$_1$, PE, and FE$_2$ states. (c) Schematic plot of the lattice potential energy surface (PES) with respect to polarizations $P_x$ and $P_z$.

We plot the DFT-calculated electronic band dispersions of the FE and PE in **Figure 2a**. Here, we assume the electric polarization *P* in the FE phase along *z* (FE$_1$, nonzero $P_z$). One sees that both phases are semiconductors with their bandgaps being 1.45 (PE) and 1.42 eV (FE), consistent with their ionic compound nature. The valence band maximum (VBM) of PE locates at the equivalent X, Y, and Z points. In the FE phase, the valley band energy at Z reduces under the depolarization field, leaving the VBM only at X (and Y). As for the lowest conduction band, one observes almost flat band dispersions along the Γ–X (or Γ–Y, Γ–Z) in the PE phase, which survives only along the Γ–Z in the FE$_1$ phase.

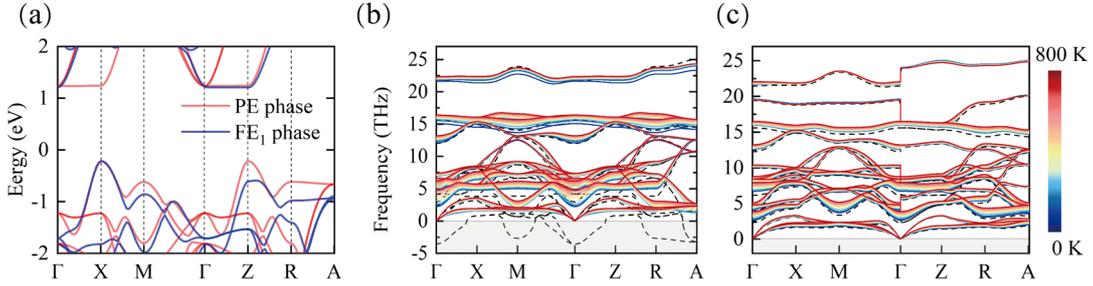

**Figure 2.** DFT-calculated fundamental properties of electrons and phonons of PbTiO$_3$. (a) Electronic band dispersion along the high symmetry ***k***-path, where blue and red curves denote the FE$_1$ and PE phase, respectively. The temperature dependent phonon spectra of the (b) PE phase and (c) FE$_1$ phase. The LO-TO splitting effect is included. The gray shaded areas are the unstable imaginary phonon frequency regime, which is denoted as negative values. The dashed curves represent the phonon dispersion according to the second harmonic approximation, while the solid curves are the finite temperature phonon structures obtained from the SCPH method.

The phonon dispersions for both PE and FE phases are plotted in **Figures 2b** and **2c**. One sees that for the PE structure, the standard second harmonic vibration results in imaginary modes (dashed curves) in the most part of the first BZ. In order to correct this and include thermal effects, we perform self-consistent phonon calculations (see

Method section). One clearly observes that the imaginary branches can be totally suppressed by including the forth order anharmonic coupling effects. Even at 0 K, it is evident that all vibration frequencies become real, showing significant renormalization effects from anharmonic motions. As the temperature increases, the frequency on each branch enhances to higher frequency. The zone-center phonon frequencies and their irreducible representation and optical activity of PbTiO$_3$ are summarized in Table 1. The irreducible representations of the twelve optical branches of PE at Γ can be decomposed as $\Gamma_{op}^{PE}(O_h) = 3T_{1u} \oplus 1T_{2u}$. As for the FE phase, there are no imaginary branches in the second harmonic framework. The irreducible representations of optical branches are $\Gamma_{op}^{FE}(C_{4v}) = 3A_1 \oplus 4E \oplus B_1$. In the vicinity of the Γ point, one observes clear frequency splitting between the longitudinal optical (LO) and transverse optical (TO) branches, ensuring the large electric dipole effects.[30, 31]

**Table 1**. The zone-center phonon frequencies (at room temperature), their corresponding irreducible representation, and the basis functions of PbTiO$_3$.

| Phase | Frequency (THz) | Irrep | Basis functions |
|---|---|---|---|
| $\Gamma_{op}^{PE}(O_h)$ | 4.6, 6.5, 16.3 | $T_{1u}$ | (x, y, z) |
|  | 7.3 | $T_{2u}$ | – |
| $\Gamma_{op}^{FE}(C_{4v})$ | 4.6, 10.1, 19.4 | $A_1$ | z, $x^2+y^2$, $z^2$ |
|  | 3.2, 7.1, 8.2, 16.2 | E | (x, y), (xz, yz) |
|  | 8.4 | $B_1$ | $x^2-y^2$ |

### B. Optical responses of PbTiO$_3$ at THz regime

Next, we calculate the frequency-dependent optical susceptibility tensor χ for these two phases, which are used to estimate the Gibbs free energy variation under linearly polarized terahertz light (LPTL) irradiation (see below and Ref. [30, 31]). The optical susceptibility originates from two sources, namely, electronic and ionic contributions.

The large electronic bandgap (> 1 eV) guarantees that the electronic contributed susceptibility almost keeps a constant in the THz regime (usually ≲ 0.4 eV). In the independent particle approximation, it can be evaluated by[34]

$$\chi_{ii}^{el} = \frac{e^2}{\epsilon_0} \int_{BZ} \frac{d^3\mathbf{k}}{(2\pi)^3} \sum_{c,v} \frac{\langle u_{v\mathbf{k}}|\nabla_{k_i}|u_{c\mathbf{k}}\rangle\langle u_{c\mathbf{k}}|\nabla_{k_i}|u_{v\mathbf{k}}\rangle}{\hbar(\omega_{c\mathbf{k}} - \omega_{v\mathbf{k}})}, \quad (1)$$

where $|u_{n\mathbf{k}}\rangle$ and $\hbar\omega_{n\mathbf{k}}$ represent the periodic part of Bloch wavefunction and its corresponding eigenenergy for band-$n$ ($c$ and $v$ refer to conduction and valence band indices, respectively) at $\mathbf{k}$. The integral is performed in the first BZ. $\epsilon_0$ is the permittivity in vacuum. We apply random phase approximation to include its nonlocal effects. Our calculations reveal that the susceptibility constant for PE is isotropic, namely, $\chi^{el}(\text{PE}) = 7.5$. The electric polarization in the FE phase gives anisotropic results, namely, $\chi_\parallel^{el}(\text{FE}) = 5.9$ (parallel to electric polarization) and $\chi_\perp^{el}(\text{FE}) = 6.6$ (normal to electric polarization).

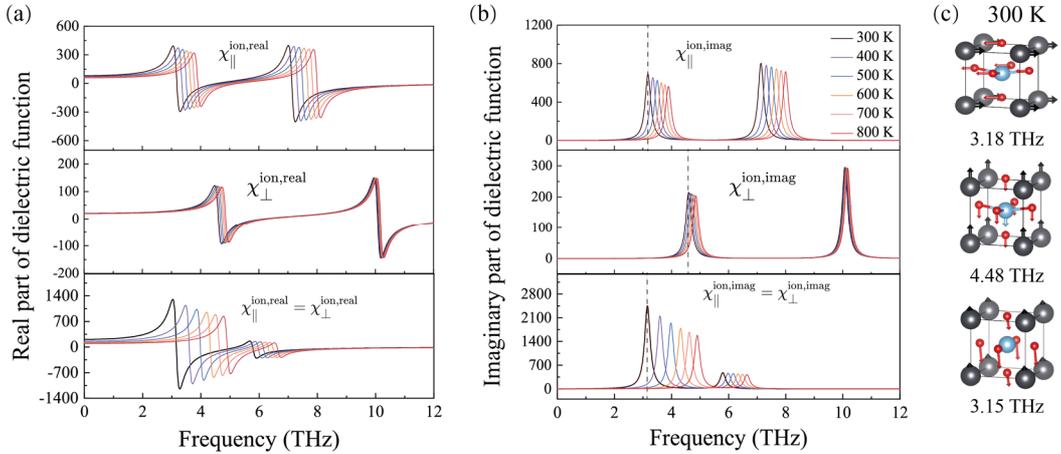

**Figure 3.** (a) Real and (b) imaginary parts of the ionic susceptibility components of PbTiO$_3$. The electric dipole for FE$_1$ is along $z$. The dashed vertical line in (b) marks the frequency of the first peak in the imaginary part of susceptibility at 300 K. (c) Schematic plot of the ionic displacement modes of the first dominant peak along each direction at room temperature.

The ionic contribution to optical susceptibility is calculated according to[30]

$$\chi_{ij}^{\text{ion}}(\omega) = \frac{1}{V} \sum_m \frac{\mathcal{Z}_{m,i}^* \mathcal{Z}_{m,j}^*}{\omega_m^2 - \left(\omega + \frac{\text{i}}{\tau^{\text{ion}}}\right)^2}, \qquad (2)$$

where $\mathcal{Z}_{m,i}^* = \sum_{\kappa,i'} z_{\kappa,ii'}^* u_{m,\kappa,i'}$ ($z_{\kappa,ii'}^*$ is Born effective charge component of ion-κ and $u_{m,\kappa,i'}$ is displacement mode) and $\omega_m$ are the Born effective charge component and eigen-frequency of vibration mode *m*, respectively. *V* is the total volume of the unit cell, $\text{i} = \sqrt{-1}$ is the unit of imaginary number, and ionic lifetime $\tau^{\text{ion}}$ (taken to be 0.12 THz) represents the ionic vibration lifetime. Note that even though such lifetime depends on phonon band index and momentum, we follow the conventional treatment to choose a universal value. It does not affect the real part of susceptibility away from the resonant frequency, which is the focus of the current study. We plot the calculated diagonal component $\chi_{ii}^{\text{ion}}(\omega)$ in **Figures 3a and 3b**. The peaks at the imaginary part of susceptibility corresponds to resonant coupling between LPTL and the IR-active modes. According to the Kramers-Kronig relation, it produces a jump in the real part of susceptibility at the resonant frequency and a finite value below such a frequency. For each of the three phases, we observe two dominant peaks at their imaginary part spectrum. Increasing temperature will shift them toward higher frequencies. **Figure 3c** schematically shows the vibration displacement modes of the first peak in the imaginary part of susceptibility at room temperature. All of them are IR-active. At a typical low frequency (e.g., 1 THz), the calculated ionic susceptibility values of the PE phase are 229.5 at room temperature. For the FE phase, the susceptibility at this low frequency $\omega_0$ is anisotropic, namely, $\chi_\parallel^{\text{ion}}(\text{FE}, \omega_0) = 87.2$ and $\chi_\perp^{\text{ion}}(\text{FE}, \omega_0) = 19.6$ at room temperature. All of the optical susceptibility components (including electronic and ionic contributions) of each phase at 1 THz frequency are listed in Table 2. Note that this frequency is away from any IR-active frequencies, hence there is no direct resonant scattering between the LPTL and the ions.

**Table 2**. The ionic and electronic contributions to susceptibility components in PbTiO$_3$ and BaTiO$_3$ (incident frequency taken to be $\omega_0$ = 1 THz at room temperature).

| Phase | PbTiO$_3$ | | BaTiO$_3$ | | | |
|---|---|---|---|---|---|---|
| | PE | FE | C | T | A | R |
| $\chi_\parallel^{\text{ion}}$ | 229.5 | 87.2 | 289.4 | 71.4 | 61.9 | 72.8 |
| $\chi_\perp^{\text{ion}}$ | 229.5 | 19.6 | 289.4 | 32.9 | 57.5 | 73.0 |
| $\chi_\parallel^{\text{el}}$ | 7.5 | 6.6 | 5.8 | 5.7 | 5.2 | 5.3 |
| $\chi_\perp^{\text{el}}$ | 7.5 | 5.9 | 5.8 | 5.1 | 5.5 | 5.3 |

### C. Optomechanics driven phase transitions in PbTiO$_3$

The structural phase transition under the off-resonant LPTL irradiation can be conducted by thermodynamic evaluations. The time-dependent alternating electric field $\vec{\mathcal{E}}(\omega, t) = \vec{E}e^{-i\omega t}$ of LPTL would reduce the Gibbs free energy (GFE) density according to $dg = -\text{Re}\langle \vec{\mathcal{P}}^* \cdot d\vec{\mathcal{E}} \rangle$, where $\langle \cdot \rangle$ indicates the time average. Note that the time-dependent electric polarization is $\vec{\mathcal{P}}(t) = \vec{P}_s + \epsilon_0 \text{Re}\overleftrightarrow{\chi}(\omega) \cdot \vec{\mathcal{E}}(\omega, t)$, where $\vec{P}_s$ is spontaneous electric polarization. Since this $\vec{P}_s$ is generally time-independent when the incident electric field strength is below the coercive field (as in the current case), its contribution to GFE would be zero after taking time average. Therefore, the GFE variation under LPTL is

$$G_{\text{LPTL}}(\vec{E}, \omega_0) = -\frac{1}{4}\epsilon_0 V \text{Re}\left[\chi_{ii}^{\text{el}}(\omega_0) + \chi_{ii}^{\text{ion}}(\omega_0)\right]E_i^2. \tag{3}$$

Here subscript *i* indicates the LPTL polarization direction. Note that the integration of $G_{\text{LPTL}}$ over electric field gives a $\frac{1}{2}$ factor, and the time average of the sinusoidal electric field gives another $\frac{1}{2}$. Hence, there is a factor of $\frac{1}{4}$ in Equation (3). Accordingly, the optical susceptibility components in different phases determine their total GFE variations under LPTL. If they show large contrast between PE, FE$_1$ and FE$_2$, LPTL would significantly change the energy profile and one could expect THz light induced thermodynamic phase transitions.

We include the LPTL contribution into the total GFE[35]

$$G(T, \vec{E}, \omega_0) = U - TS + pV + G_{\text{LPTL}}(\vec{E}, \omega_0). \qquad (4)$$

Here, $U$ is the total internal energy, and $S$ is the entropy. We only consider the ion contributed entropy in our discussion, as the electron contributed entropy is significantly small for intrinsic semiconductors with bandgap larger than 1 eV. At the pressure-free condition, the third term is safely ignored. Before the LPTL is irradiated, our numerical results yield a FE to PE transition at ~ 720 K, closing to the experimental observations.[36] This validates the computational approaches with the anharmonic phonon effects.

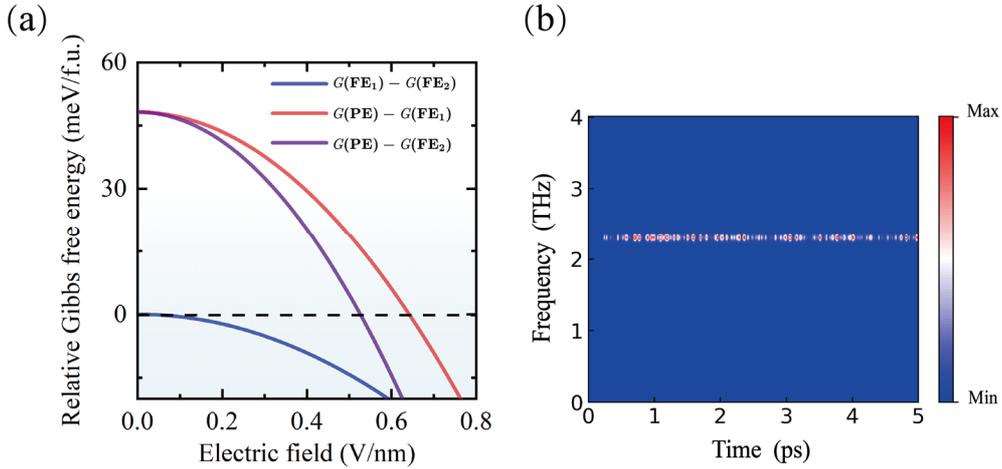

**Figure 4.** (a) Relative GFE of PbTiO$_3$ as a function of LPTL electric field magnitude at room temperature. (b) Non-resonant phonon response under alternating electric field of FE PbTiO$_3$. Only a Raman-active mode appears here.

Without loss of generality, we assume that the light travels along the *z*-direction, hence light electric field is polarized in the *xy*-plane (we take *i* = *x* or *y* here). According to Equation (4), we now calculate the GFE difference for phase transitions between the FE$_1$/FE$_2$ (Goldstone mode) and FE/PE (Higgs mode), as shown in **Figure 4a**. One sees that before light is introduced (electric field *E* = 0 V/nm), the GFE difference between FE and PE is 48.2 meV per formula unit (f.u.) at room temperature. The two FE phases are energetically degenerate. When we apply a LPTL (at frequency $\omega_0$) with its polarization along *x*, the FE$_1$/FE$_2$ degeneracy lifts due to anisotropic optical responses, namely, different $\chi_\perp$ and $\chi_\parallel$. Hence, we would have a phase transition from FE$_2$ (smaller $\chi_\perp$) to FE$_1$ (larger $\chi_\parallel$). The GFE difference is 14.3 meV/f.u. under 0.5 V/nm electric field, sufficiently large to be distinguished (Goldstone mode). Similarly, as the susceptibility $\chi$ of PE is much larger than that of the FE. Above a critical electric field of 0.6 – 0.7 V/nm (depending on the initial structure being FE$_1$ or FE$_2$), the LPTL drives the PE structure to be stable thermodynamically. This implies a FE to PE phase transformation (Higgs mode). These phase transitions do not require any bond breaking and reformation, and only involve collective and coherence atomic displacive motions. Hence, they would occur in a fast kinetics and do not generate significant heat as in the resonant light absorption mechanism.

Since the 1 THz light is off-resonant to excite either an electron-hole pair in electronic band structure or an IR-active mode in the phonon vibration, one may wonder how the light-matter interaction occurs. According to our previous work[32, 33], we use a simple expansion on $\chi$ to propose that it belongs to a second order nonlinear optical process, which stimulate a Raman-active mode. In this work, we perform an alternating electric field dressed *ab initio* molecular dynamics to illustrate this conclusion. We apply the modern theory of polarization to evaluate the time-dependent *P* in each step, and multiply it with time-dependent electric field $\mathcal{E}(\omega_0, t)$ (see Methods for more details). We then project the ionic vibrational offset into normal coordinates of each phonon calculated by density functional perturbation theory (DFPT). **Figure 4b** shows that applying an *x*-polarized LPTL onto the FE$_1$ PbTiO$_3$ (under an alternating field with a magnitude of 0.1 V/nm) could induce a non-resonant excitation of phonon mode at 2.3 THz (second order harmonic calculation), which corresponds to the 3.2 THz from the anharmonic SCPH calculation (Table 1). This is a Raman-active mode (irreducible representation of *E*), consistent with the nonlinear optical process nature.

***D. THz induced phase transitions in BaTiO$_3$***

In order to show the ubiquity of this optomechanical effect, we take another prototypical perovskite, barium titanate (BaTiO$_3$), which exhibits more isomeric phases. The BaTiO$_3$ undergoes displacive structural phase transitions from a high temperature centrosymmetric phase ($Pm\bar{3}m$, denoted as C) into tetragonal ($P4mm$, denoted as T) at 409 K, which breaks inversion symmetry by polar displacements of Ti and O atoms. Below 278 K, it transits into an orthorhombic phase ($Amm2$, A) and then a rhombohedral phase ($R3m$, R) at 183 K.[37] We perform DFT calculations and the self-consistent phonon calculations to evaluate phonon spectra of these four phases, as shown in **Figure 5**. Similar as above, we denote two orientational variants of T phase as T$_1$ phase (with electric dipole along $x$) and T$_2$ phase (with electric dipole along $z$). Note that if only second harmonic approximation is applied, only the R phase shows no imaginary frequencies (dashed curves). In the self-consistent phonon calculation framework, we find all the imaginary phonon branches disappear in these four phases, even at 0 K. This clearly shows the significance of inclusion the high order anharmonic interactions.

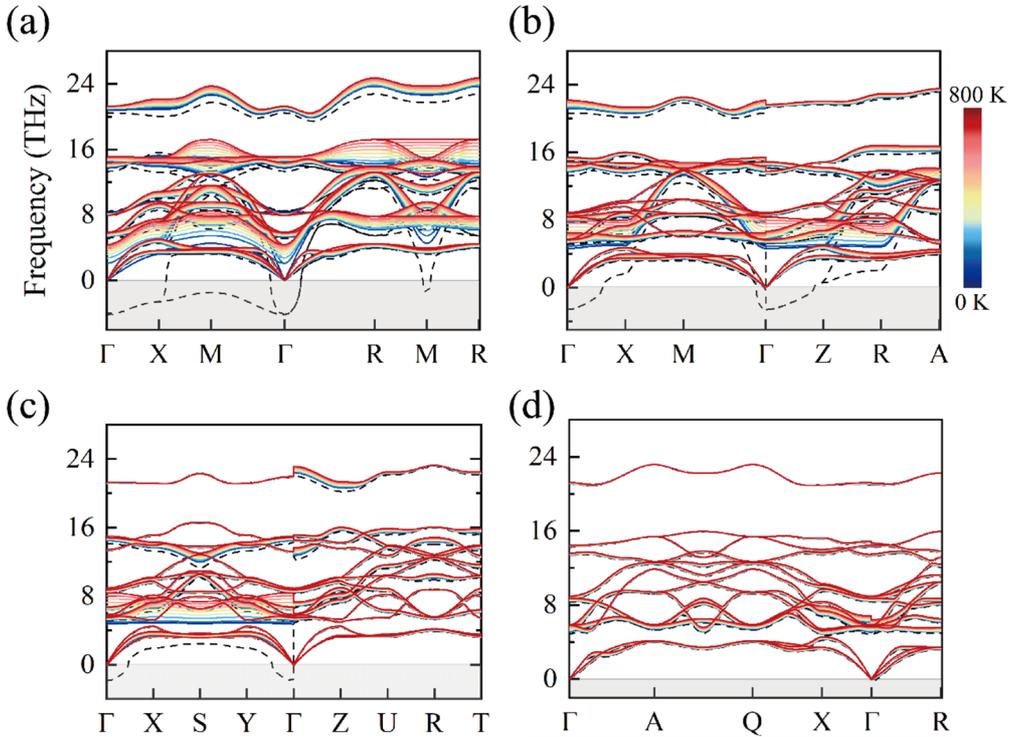

**Figure 5.** Temperature dependent phonon spectra of BaTiO$_3$ in the (a) C, (b) T$_1$, (c) A, and (d) R phase. The dashed curves represent the phonon dispersion according to the second harmonic approximation, while the solid curves are from the SCPH scheme.

We then compute the optical responses of each phase in BaTiO$_3$, which are listed in Table 2. According to Equation (4), we estimate the changes in GFE of different phases under 1 THz illumination (at room temperature, with its electric field polarized along $x$), which is depicted in **Figure 6**. It suggests that above a critical electric field strength of ~0.3 V/nm, the T phase would transit into the C phase, which is high symmetric and have vanished spontaneous electric polarization. This thermodynamic result is counterintuitive to the general physical picture that adding an electric field induces dipole moment (as in the conventional $E - P$ hysteresis). The underlining reason is that our optomechanical approach contains second order nonlinear optical response, while the $E - P$ hysteresis is only the first order response. In addition, one notes that the C phase usually emerges under high temperature, but here we show that it could appear at room temperature under THz irradiation, being a hidden phase.

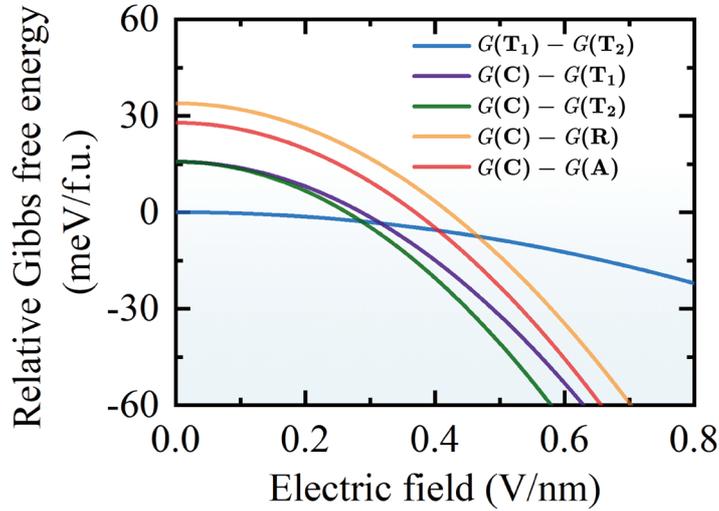

**Figure 6.** Relative Gibbs free energy of BaTiO$_3$ as a function of LPTL electric field magnitude under room temperature. The symbols $T_1$ ($T_2$), C, R, and A denote tetrahedral, cubic, rhombohedral, and orthorhombic phases, respectively. Here $T_1$ and $T_2$ refer to that the electric polar direction is parallel and perpendicular to the light electric field direction, respectively.

**III. Conclusion**

In conclusion, we demonstrate that intermediate THz could drive phase transitions in perovskite oxides such as PbTiO3 and BaTiO3, controlling the emergence of ferroelectric polarization and its direction. We show that anharmonic ionic interactions strongly affect the phonon spectrum and the optical susceptibility, which significantly varies as a function of temperature. Using electric field dressed *ab initio* molecular dynamics simulations, we explicitly show that the off-resonant THz irradiation would evoke the Raman phonon vibration (rather than the resonant IR-active modes). The orientational variant transition provides a theoretical explanation for the recent experimental observations.[13] This optomechanics induced phase transitions could be easily generalized to other perovskite systems and ferroelectric materials, serving an applicable platform for information storage and in-memory computing devices.

## IV. Methods

Our DFT calculations are performed in the Vienna *ab initio* simulation package (VASP)[38] that treats the exchange-correlation functional using the generalized gradient approximation (GGA) method in the solid state Perdew-Burke-Ernzerhof (PBEsol) form.[39] Projector augmented-wave (PAW) method is used to treat the core electrons, yielding an accuracy compared with all-electron approaches.[40, 41] The valence electrons (Ti: $4s^23d^2$, O: $2s^22p^4$, Pb: $5d^{10}6s^26p^2$, and Ba: $5s^25p^66s^2$) are described by a planewave basis set with a kinetic cutoff energy of 500 and 550 eV for PbTiO3 and BaTiO3, respectively. The first Brillouin zone (BZ) is represented by Monkhorst-Pack $k$-mesh grids of (12×12×12).[42] The total energy and force convergence criteria are set as $1 \times 10^{-8}$ eV and $1 \times 10^{-8}$ eV Å$^{-1}$, respectively. The convergence of these computational parameters has been carefully tested. To include the anharmonic effect in the phonon dispersion, we adopt the self-consistent approach (SCPH scheme[43]) with microscopic anharmonic force constants extracted from density functional calculations using the least absolute shrinkage and selection operator technique.[44] This method has been successful in reproducing the finite-temperature phonon dispersion of various polar solids, which exhibits well-consistent results with experimental observations. The simulation supercell is composed by (2 × 2 × 2) unit cells containing 40 atoms, which are used to generate force constants from randomly selected structures during *ab initio* molecular dynamics (AIMD) simulations.[45] Non-analytic correction is included by the Ewald method and the Born effective charge components are evaluated according to the density functional perturbation theory. The Brillouin zone sampling of vibrational free energy from SCPH approach, $F_{\text{vib}}^{(\text{SCPH})}(V,T)$, is computed based on (20 × 20 × 20) *q*-mesh grids in the first BZ. The trajectory of atomic motion under an alternating electric field is simulated by combining electric fields in the AIMD method. The temperature of 300 K and the microcanonical ensemble were selected as the simulation conditions. The zero-temperature equilibrium structure is used as the initial structure to ensure that no phonons are activated before simulation. Density functional perturbation theory calculation results are used as reference basis for the projection of molecular dynamics trajectories into phonon modes.

**Acknowledgments.** The authors thank Dr. Yongliang Shi, Tingwei Li, and Penghu Du for valuable discussions on the simulation procedure, and acknowledge Chuanwei Fan for the computational technical supports. This work was supported by National Natural Science Foundation of China (NSFC) under Grant Nos. 21903063 and 11974270. The calculations are performed on the HPC platform of Xi'an Jiaotong University.

**Author Contribution.** J.Z. conceived the concept, C.Z. performed the calculation. Both authors analyzed the data and wrote the manuscript.

**Data Availability.** The data that support the findings of this study are included in this article and are available from the corresponding author upon reasonable request.